\documentclass[showpacs,aps,graphicx]{revtex4}
\usepackage{graphicx}

\begin{document}
\title{Complete analysis for arbitrary concatenated Greenberger-Horne-Zeilinger  state assisted with photonic Faraday rotation}

\author{Lan Zhou$^{1,2}$, Shi-Pu Gu$^{3}$, Xing-Fu Wang$^{1}$, Yu-Bo Sheng,$^{2}$\footnote{Email address:
shengyb@njupt.edu.cn}  }
\address{
 $^1$ College of Mathematics \& Physics, Nanjing University of Posts and Telecommunications, Nanjing,
210003, China\\
 $^2$Key Lab of Broadband Wireless Communication and Sensor Network
 Technology, Nanjing University of Posts and Telecommunications, Ministry of
 Education, Nanjing, 210003, China\\
 $^3$College of Electronic Science and Engineering, Nanjing University of Posts and Telecommunications,
Nanjing, 210003, China\\}

\begin{abstract}
The concatenated Greenberger-Horne-Zeilinger (C-GHZ) state has great  potential application in the future quantum network, for it is robust to the decoherence in a noisy environment. In the paper, we propose a complete C-GHZ state analysis protocol with the help of some auxiliary single atoms trapped in the low-quality cavities. In the protocol, we essentially make the parity check for the photonic states based on the photonic Faraday rotation effect, and complete the analysis task combined with the Hadamard operation and single qubit measurement.  The success probability of our protocol can reach 100\% in principle, and the number of physical qubit encoded in each logic qubit does not affect the analysis. Our analysis protocol may have its practical application in future
long-distance quantum communication.
\end{abstract}
 \maketitle

\section{Introduction}

Entanglement is the key resource of quantum information processing (QIP). Entangled qubits are required for many important branches of QIP, such as quantum teleportation \cite{t1}, quantum key distribution (QKD) \cite{QKD1}, quantum secure direct communication (QSDC) \cite{QSDC1,QSDC2}, and quantum repeaters \cite{repeater}. Recently, the multi-particle systems attract
more and more attentions. For instance, the Greenberger-Horne-Zeilinger (GHZ) state is one of the most important resources in QIP. Due to the large information capacity, GHZ states play an important role in
the foundations of quantum mechanics measurement theory
and quantum communication \cite{GHZ1,GHZ2,GHZ3,GHZ4}. In practical applications, the ideal quantum states are the maximally entangled states. However, due to the environmental noise, the decoherence problem is inevitable in practical applications. The decoherence greatly limits the building of high-quality quantum channel, even more, it may cause the quantum communication insecure. For dealing with the decoherence problem, people proposed large number of approaches, such as entanglement purification \cite{EPP1,EPP2,EPP3,EPP4,EPP5,EPP6,EPP7,EPP8,EPP9,EPP10}, entanglement concentration \cite{ECP1,ECP2,ECP3,ECP4,ECP5,ECP6,ECP7}, and entanglement amplification \cite{zhouNLA,zhoujosa2,amplification}. Recently, a new kind of multi-particle quantum state, which is called the concatenated Greenberger-Horne-Zeilinger (C-GHZ) state has attracted high attention \cite{logic,cghz2,logic1,logic2,logic3,logic4}. It is also called the macroscopic Schr\"{o}dinger's cat superposed state. For the common quantum states, people usually encode quantum qubit in physical qubit directly, while for the C-GHZ state, the parties encode many physical qubits in a logic qubit. The typical C-GHZ state can be written as
\begin{eqnarray}
|\Phi_{1}^{\pm}\rangle_{N,m}=\frac{1}{\sqrt{2}}(|GHZ^{+}_{m}\rangle^{\otimes N} \pm |GHZ^{-}_{m}\rangle^{\otimes N}).\label{logic}
\end{eqnarray}
 Here, $|GHZ^{\pm}_{m}\rangle=\frac{1}{\sqrt{2}}(|0\rangle^{\otimes m}\pm|1\rangle^{\otimes m})$. $N$ and $m$ are the logic qubit number and physical qubit number in each logic qubit, respectively. This C-GHZ state shows similar features as
the common GHZ state. However, comparing with common GHZ state, the C-GHZ state has a highly attractive feature, that is, it is robust to the decoherence in a noisy environment \cite{logic,logic1}. Due to the robust feature, the C-GHZ state has great application potential in the future long-distance quantum communication. In 2014, Lu \emph{et al.} demonstrated the first experiment to prepare the C-GHZ state with $M=2$ and $N=3$ in an optical system. They also verified that the C-GHZ state can tolerate more  bit-flip and phase shift noise
 than polarized GHZ state. Therefore, the C-GHZ state is useful for large-scale fibre-based quantum networks and
multipartite QKD schemes \cite{logic1}.

 The quantum state analysis, which is the discrimination between the maximally entangled quantum states is quite important in various applications. The most common quantum state analysis is called the Bell-state analysis (BSA), which is for the two-particle entangled system. There are usually three different kinds of methods to realize the BSA. The first one is totally in linear optics \cite{BSA1,BSA2}. However, the success probability of the BSA approaches with only linear optical elements can only reach 50\%, so that the first kind of BSA approaches can not perform complete BSA \cite{BSA1,BSA2}. The second kind of methods still requires the linear optical elements but resorts to the hyperentanglement \cite{hyper1,hyper2,hyper3,hyper4}. For example, in 2003, Walborn \emph{et al.}
once proposed a hyperentanglement-assisted BSA approach. In their approach,
the hyperentangled state is prepared in polarization and momentum degrees of freedom. They can realize the complete BSA for both the momentum and polarization entangled Bell-states \cite{hyper1}. The third kind of methods adopt the nonlinear optical elements, such as the cross-Kerr nonlinearity and the quantum-dot system \cite{kerr1,kerr2,dot1,dot2,QND1} to realize the complete BSA. For example, some groups adopt the cross-Kerr nonlinearity to construct the complete parity-check measurement (PCM) gate, which can distinguish the even parity states $|H\rangle|H\rangle$ and $|V\rangle|V\rangle$ from the odd parity states
$|H\rangle|V\rangle$ and $|V\rangle|H\rangle$ \cite{kerr1,QND1}. Although the analysis for GHZ states has been widely discussed \cite{kerr2,GHZA1,GHZA2,GHZA3,GHZA4,GHZA5,GHZA7}, most analysis protocols can not deal with the C-GHZ states. Recently, Sheng and Zhou proposed two complete BSA protocols for the C-GHZ state with the help of the controlled-not (CNOT) gate and the cross-Kerr nonlinearity, respectively \cite{shengepl,shengsr}. Unfortunately, the CNOT gate and the cross-Kerr nonlinearity are difficult to realize in current experimental condition, which limits the application of the two analysis protocols. Lee \emph{et al.} proposed a partially BSA protocol for another type of logic-qubit entanglement in the linear optics \cite{BSA}. 

On the other hand, the cavity quantum electrodynamics (QED) is a promising
platform for performing quantum information tasks, due to the controllable
interaction between atoms and photons. In 2009, the group of An successfully implemented QIP based on the photonic Faraday rotation \cite{cavity}. This method works in the low-quality (Q) cavities and only involves
the virtual excitation of atoms. Therefore, it is insensitive to both
the cavity decay and the atomic spontaneous emission. Following this scheme, various works based on the photonic Faraday rotation effect in the low-Q cavity
have been presented, such as quantum logic gate \cite{gate}, QIP in decoherence-free subspace \cite{gate1},
quantum teleportation \cite{cavity1}, and entanglement detection \cite{zhouatom,zhouentropy}. Recently, with the help of Faraday rotation, Wei and Deng designed some compact quantum gates \cite{wei1,wei2}. Their works proved that the
universal quantum computation can be realized. Recently, we proposed a complete logic Bell-state analysis (LBSA) protocol with the help of the photonic Faraday rotation in low-Q cavity \cite{zhoupra2}. Actually, the logic Bell-state is the special case of the C-GHZ state with $N=2$. In this paper, we will put forward a complete analysis protocol for the C-GHZ state with arbitrary $N$ and $m$ based on the photonic Faraday rotation in low-Q cavity. Due to the attractive possible applications of the C-GHZ state, our analysis may be useful in the future QIP field.

 This paper is organized as follows: In section 2, we will introduce the basic principle of the photonic Faraday rotation.
In section 3, we will describe our complete analysis protocol for arbitrary C-GHZ state in detail. In section 4, we make
 a discussion and conclusion.

\section{Basic principle of the photonic Faraday rotation}
\begin{figure}[!h]
\begin{center}
\includegraphics[width=7cm,angle=0]{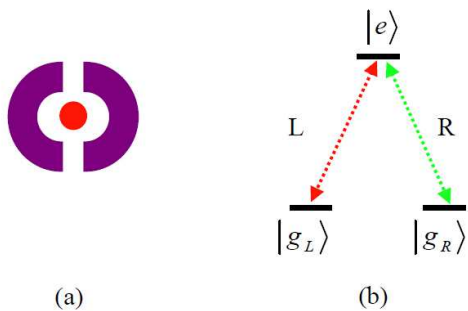}
\caption{The atomic structure in a low-Q cavity field. A three-level atom is trapped in a low-Q cavity (a). It has an excited state $|e\rangle$ and two degenerate ground states $|g_{L}\rangle$ and $|g_{R}\rangle$. The transition between $|g_{L}\rangle$ and $|e\rangle$ is assisted with a left-circularly polarized photon ($|L\rangle$), and the transition between $|g_{R}\rangle$ and $|e\rangle$ is assisted with a right-circularly polarized photon ($|R\rangle$), respectively. (b).}
\end{center}
\end{figure}

Our analysis protocol is based on the photonic Faraday rotation in low-Q cavity. In this way, we first introduce its basic principle briefly. As shown in Fig. 1, a three-level atom is trapped in the one-side low-Q cavity. The states $|g_{L}\rangle$ and $|g_{R}\rangle$ represent the two Zeeman sublevels of its degenerate ground state, and $|e\rangle$ represents its excited state. A single photon pulse with frequency $\omega_{p}$ enters the cavity and reacts with the three-level atom. The transition between $|g_{L}\rangle$ and $|e\rangle$ is assisted with a left-circularly polarized photon ($|L\rangle$), and the transition between $|g_{R}\rangle$ and $|e\rangle$ is assisted with a right-circularly polarized
photon ($|R\rangle$), respectively.

Based on the researches of Res. \cite{cavity,cavity1,cteleportationqip,Faraday}, by solving the Langevin equations of
motion for cavity and atomic lowering operators analytically, we can obtain the general expression of the reflection coefficient of the atom-cavity
system in the form of
\begin{eqnarray}
r(\omega_{p})\equiv\frac{a_{out,j(t)}}{a_{in,j(t)}}=\frac{[i(\omega_{c}-\omega_{p})-\frac{\kappa}{2}][i(\omega_{0}-\omega_{p})+\frac{\gamma}{2}]+\lambda^{2}}
{[i(\omega_{c}-\omega_{p})+\frac{\kappa}{2}][i(\omega_{0}-\omega_{p})+\frac{\gamma}{2}]+\lambda^{2}}.\label{r}
\end{eqnarray}
Here, $a_{in}(t)$ and $a_{out}(t)$ are the cavity input operator and cavity output operator, respectively. $\kappa$ and $\gamma$ are the cavity damping rate and atomic decay rate, respectively. $\omega_{c}$, and $\omega_{0}$ are the frequency of the cavity and the atom, respectively, and $\lambda$ is the atom-cavity coupling strength. From Eq. (\ref{r}), if the atom uncouples to the cavity, which makes $\lambda=0$, we can simplify $r(\omega_{p})$ to
\begin{eqnarray}
r_{0}(\omega_{p})=\frac{i(\omega_{c}-\omega_{p})-\frac{\kappa}{2}}{i(\omega_{c}-\omega_{p})+\frac{\kappa}{2}}.\label{r0}
\end{eqnarray}
Eq. (\ref{r0}) can be written as a pure phase shift $r_{0}(\omega_{p})=e^{i\phi_{0}}$. On the other hand, in the interaction process, as the photon experiences an extremely weak absorption, we can consider that the output reflected photon only experiences a pure phase shift without any absorption. In this way, the expression of $r(\omega_{p})$ can be simplified to $r(\omega_{p})\simeq e^{i\phi}$. Therefore, if the photon pulse takes action, the output photon state will convert to $|\varphi_{out}\rangle=r(\omega_{p})|L (R)\rangle\simeq e^{i\phi}|L (R)\rangle$, otherwise, the single-photon would only sense the empty cavity, and the output photon state will convert to $|\varphi_{out}\rangle=r_{0}(\omega_{p})|L (R)\rangle= e^{i\phi_{0}}|L (R)\rangle$.

In this way, for an input single-photon state $|\varphi_{in}\rangle=\frac{1}{\sqrt{2}}(|L\rangle+|R\rangle)$, if the initial atom state is $|g_{L}\rangle$, the output photon state can evolve to
\begin{eqnarray}
|\varphi_{out}\rangle_{-}=\frac{1}{\sqrt{2}}(e^{i\phi}|L\rangle+e^{i\phi_{0}}|R\rangle),\label{L}
\end{eqnarray}
while if the initial atom state is $|g_{R}\rangle$, the output photon state will evolve to
\begin{eqnarray}
|\varphi_{out}\rangle_{+}=\frac{1}{\sqrt{2}}(e^{i \phi_{0}}|L\rangle+e^{i \phi}|R\rangle).\label{R}
\end{eqnarray}
Based on Eq. (\ref{L}) and Eq. (\ref{R}), the angle $\Theta^{-}_{F}=\phi_{0}-\phi$ or $\Theta^{+}_{F}=\phi-\phi_{0}$ is defined as the photonic Faraday rotation.

In  Eq. (\ref{r}) and Eq. (\ref{r0}), it can be found that in a certain case, i.e., $\omega_{0}=\omega_{c}$, $\omega_{p}=\omega_{c}-\frac{\kappa}{2}$, and $\lambda=\frac{\kappa}{2}$, we can obtain $\phi=\pi$ and $\phi_{0}=\frac{\pi}{2}$, so that the relation between the input and output photonic state can be written  as \cite{zhouatom,zhouentropy,Pengpra}
\begin{eqnarray}
&&|L\rangle|g_{L}\rangle\rightarrow -|L\rangle|g_{L}\rangle,\qquad |R\rangle|g_{L}\rangle\rightarrow i|R\rangle|g_{L}\rangle,\nonumber\\
&&|L\rangle|g_{R}\rangle\rightarrow i|L\rangle|g_{R}\rangle,\qquad |R\rangle|g_{R}\rangle\rightarrow -|R\rangle|g_{R}\rangle.\label{relation}
\end{eqnarray}

\section{The complete C-GHZ state analysis with $N=3$}
\begin{figure}[!h]
\begin{center}
\includegraphics[width=14cm,angle=0]{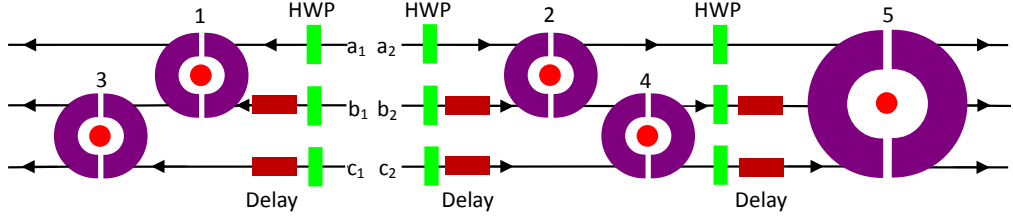}
\caption{The schematic drawing of the complete analysis protocol for the C-GHZ states with $N=3$ and $m=2$. The parties make five three-level atoms "1", "2", "3", "4", and "5"  trap in five low-Q cavities, respectively. The first four atoms are prepared in the same state of $\frac{1}{\sqrt{2}}(|g_{L}\rangle+|g_{R}\rangle)$, and the atom "5" is prepared in $|g_{L}\rangle$. HWP represents the half-wave plate, which makes $|L\rangle\rightarrow\frac{1}{\sqrt{2}}(|L\rangle+|R\rangle)$ and $|R\rangle\rightarrow\frac{1}{\sqrt{2}}(|L\rangle-|R\rangle)$. The "Delay" represents the time-delay setup, which is adopted to ensure each cavity only contains one photon at a time. After the photon-atom interaction, the photons are reflected and exit the cavity.}
\end{center}
\end{figure}
In the section, we introduce the complete analysis protocol for the C-GHZ states under a simply case, say $N=3$. For simplicity, we first suppose only two physical qubits encoded in each logic qubit, that is, $m=2$. Under this case, we can write eight C-GHZ states as
\begin{eqnarray}
|\Phi^{\pm}_{1}\rangle_{3,2}&=&\frac{1}{\sqrt{2}}(|\phi^{+}\rangle_{A}|\phi^{+}\rangle_{B}|\phi^{+}\rangle_{C}\pm|\phi^{-}\rangle_{A}|\phi^{-}\rangle_{B}|\phi^{-}\rangle_{C}),\nonumber\\
|\Phi^{\pm}_{2}\rangle_{3,2}&=&\frac{1}{\sqrt{2}}(|\phi^{-}\rangle_{A}|\phi^{+}\rangle_{B}|\phi^{+}\rangle_{C}\pm|\phi^{+}\rangle_{A}|\phi^{-}\rangle_{B}|\phi^{-}\rangle_{C}),\nonumber\\
|\Phi^{\pm}_{3}\rangle_{3,2}&=&\frac{1}{\sqrt{2}}(|\phi^{+}\rangle_{A}|\phi^{-}\rangle_{B}|\phi^{+}\rangle_{C}\pm|\phi^{-}\rangle_{A}|\phi^{+}\rangle_{B}|\phi^{-}\rangle_{C}),\nonumber\\
|\Phi^{\pm}_{4}\rangle_{3,2}&=&\frac{1}{\sqrt{2}}(|\phi^{+}\rangle_{A}|\phi^{+}\rangle_{B}|\phi^{-}\rangle_{C}\pm|\phi^{-}\rangle_{A}|\phi^{-}\rangle_{B}|\phi^{+}\rangle_{C}),\label{multi1}
\end{eqnarray}
where \begin{eqnarray}
|\phi^{\pm}\rangle=\frac{1}{\sqrt{2}}(|L\rangle|L\rangle\pm|R\rangle|R\rangle),\nonumber\\
|\psi^{\pm}\rangle=\frac{1}{\sqrt{2}}(|L\rangle|R\rangle\pm|R\rangle|L\rangle).
\end{eqnarray}

The schematic drawing of our complete analysis protocol is shown in Fig. 2. The protocol includes two steps. In the first step, we first make each of the photons pass through a half-wave plate (HWP). In essence, the HWP plays the role of the Hadamard operation, which makes  $|L\rangle\rightarrow\frac{1}{\sqrt{2}}(|L\rangle+|R\rangle)$ and $|R\rangle\rightarrow\frac{1}{\sqrt{2}}(|L\rangle-|R\rangle)$. After the HWP, $|\phi^{+}\rangle$ will not change, while $|\phi^{-}\rangle$ will change to $|\psi^{+}\rangle$. In this way, after the HWPs, the eight C-GHZ states in Eq. (\ref{multi1}) will evolve to
\begin{eqnarray}
|\Phi^{\pm}_{1}\rangle_{3,2}&=&\frac{1}{\sqrt{2}}(|\phi^{+}\rangle_{A}|\phi^{+}\rangle_{B}|\phi^{+}\rangle_{C}\pm|\psi^{+}\rangle_{A}|\psi^{+}\rangle_{B}|\psi^{+}\rangle_{C}),\nonumber\\
|\Phi^{\pm}_{2}\rangle_{3,2}&=&\frac{1}{\sqrt{2}}(|\psi^{+}\rangle_{A}|\phi^{+}\rangle_{B}|\phi^{+}\rangle_{C}\pm|\phi^{+}\rangle_{A}|\psi^{+}\rangle_{B}|\psi^{+}\rangle_{C}),\nonumber\\
|\Phi^{\pm}_{3}\rangle_{3,2}&=&\frac{1}{\sqrt{2}}(|\phi^{+}\rangle_{A}|\psi^{+}\rangle_{B}|\phi^{+}\rangle_{C}\pm|\psi^{+}\rangle_{A}|\phi^{+}\rangle_{B}|\psi^{+}\rangle_{C}),\nonumber\\
|\Phi^{\pm}_{4}\rangle_{3,2}&=&\frac{1}{\sqrt{2}}(|\phi^{+}\rangle_{A}|\phi^{+}\rangle_{B}|\psi^{+}\rangle_{C}\pm|\psi^{+}\rangle_{A}|\psi^{+}\rangle_{B}|\phi^{+}\rangle_{C}).\label{multi2}
\end{eqnarray}

The parties make four three-level atoms, here named atom "1", "2", "3", and "4" trap in four low-Q cavities, respectively. The four atoms are prepared in the same states as $|\Omega_{1}\rangle=|\Omega_{2}\rangle=|\Omega_{3}\rangle=|\Omega_{4}\rangle=\frac{1}{\sqrt{2}}(|g_{L}\rangle+|g_{R}\rangle)$. Then, the parties make the photons in $a_{1}b_{1}$ and $a_{2}b_{2}$ spatial modes pass through the cavities and interact with the atoms "1" and $2$, respectively. After the photons in $b_{1}$ and $b_{2}$ modes exiting the cavities, the parties make the photons in $b_{1}c_{1}$ and $b_{2}c_{2}$ modes enter another two cavities and interact with the atoms "3" and "4", respectively. It is noticed that we should ensure that only a photon interacts with the atom at a time. In this way, we adopt the setup "Delay" to exactly control the time of the photon entering the cavity. For example, as shown in Fig. 2, we first let the photon in $a_{1}$ mode interact with the atom "1". After the photon is reflected and exits the cavity, we let the photon in $b_{1}$ mode enter the cavity.

 If the initial C-GHZ state is $|\Phi^{\pm}_{1}\rangle_{3,2}$, the photon state combined with the four atom states can be written as
\begin{eqnarray}
&&|\Phi^{\pm}_{1}\rangle_{3,2}\otimes|\Omega_{1}\rangle\otimes|\Omega_{2}\rangle\otimes|\Omega_{3}\rangle\otimes|\Omega_{4}\rangle\nonumber\\
&=&\frac{1}{\sqrt{2}}[\frac{1}{\sqrt{2}}(|LL\rangle_{a1a2}+|RR\rangle_{a1a2})\frac{1}{\sqrt{2}}(|LL\rangle_{b1b2}+|RR\rangle_{b1b2})\frac{1}{\sqrt{2}}(|LL\rangle_{c1c2}+|RR\rangle_{c1c2})\nonumber\\
&\pm&\frac{1}{\sqrt{2}}(|LR\rangle_{a1a2}+|RL\rangle_{a1a2})\frac{1}{\sqrt{2}}(|LR\rangle_{b1b2}+|RL\rangle_{b1b2})\frac{1}{\sqrt{2}}(|LR\rangle_{c1c2}+|RL\rangle_{c1c2})]\nonumber\\
&\otimes&\frac{1}{\sqrt{2}}(|g_{L}\rangle_{1}+|g_{R}\rangle_{1})\otimes\frac{1}{\sqrt{2}}(|g_{L}\rangle_{2}+|g_{R}\rangle_{2})\otimes\frac{1}{\sqrt{2}}(|g_{L}\rangle_{3}+|g_{R}\rangle_{3})
\otimes\frac{1}{\sqrt{2}}(|g_{L}\rangle_{4}+|g_{R}\rangle_{4})\nonumber\\
&=&\frac{1}{4}[(|L\rangle_{a1}|L\rangle_{a2}|L\rangle_{b1}|L\rangle_{b2}|L\rangle_{c1}|L\rangle_{c2}+|L\rangle_{a1}|L\rangle_{a2}|L\rangle_{b1}|L\rangle_{b2}|R\rangle_{c1}|R\rangle_{c2}
+|L\rangle_{a1}|L\rangle_{a2}|R\rangle_{b1}|R\rangle_{b2}|L\rangle_{c1}|L\rangle_{c2}\nonumber\\
&+&|L\rangle_{a1}|L\rangle_{a2}|R\rangle_{b1}|R\rangle_{b2}|R\rangle_{c1}|R\rangle_{c2}+|R\rangle_{a1}|R\rangle_{a2}|L\rangle_{b1}|L\rangle_{b2}|L\rangle_{c1}|L\rangle_{c2}
+|R\rangle_{a1}|R\rangle_{a2}|L\rangle_{b1}|L\rangle_{b2}|R\rangle_{c1}|R\rangle_{c2}\nonumber\\
&+&|R\rangle_{a1}|R\rangle_{a2}|R\rangle_{b1}|R\rangle_{b2}|L\rangle_{c1}|L\rangle_{c2}+|R\rangle_{a1}|R\rangle_{a2}|R\rangle_{b1}|R\rangle_{b2}|R\rangle_{c1}|R\rangle_{c2})
\pm(|L\rangle_{a1}|R\rangle_{a2}|L\rangle_{b1}|R\rangle_{b2}|L\rangle_{c1}|R\rangle_{c2}\nonumber\\
&+&|L\rangle_{a1}|R\rangle_{a2}|L\rangle_{b1}|R\rangle_{b2}|R\rangle_{c1}|L\rangle_{c2}+|L\rangle_{a1}|R\rangle_{a2}|R\rangle_{b1}|L\rangle_{b2}|L\rangle_{c1}|R\rangle_{c2}
+|L\rangle_{a1}|R\rangle_{a2}|R\rangle_{b1}|L\rangle_{b2}|R\rangle_{c1}|L\rangle_{c2}\nonumber\\
&+&|R\rangle_{a1}|L\rangle_{a2}|L\rangle_{b1}|R\rangle_{b2}|L\rangle_{c1}|R\rangle_{c2}+|R\rangle_{a1}|L\rangle_{a2}|L\rangle_{b1}|R\rangle_{b2}|R\rangle_{c1}|L\rangle_{c2}
+|R\rangle_{a1}|L\rangle_{a2}|R\rangle_{b1}|L\rangle_{b2}|L\rangle_{c1}|R\rangle_{c2}\nonumber\\
&+&|R\rangle_{a1}|L\rangle_{a2}|R\rangle_{b1}|L\rangle_{b2}|R\rangle_{c1}|L\rangle_{c2})]\otimes\frac{1}{\sqrt{2}}(|g_{L}\rangle_{1}+|g_{R}\rangle_{1})
\otimes\frac{1}{\sqrt{2}}(|g_{L}\rangle_{2}+|g_{R}\rangle_{2})\nonumber\\
&\otimes&\frac{1}{\sqrt{2}}(|g_{L}\rangle_{3}+|g_{R}\rangle_{3})
\otimes\frac{1}{\sqrt{2}}(|g_{L}\rangle_{4}+|g_{R}\rangle_{4}).\label{0}
\end{eqnarray}
According to the input-output relation in Eq. (\ref{relation}), after the cavities, the states in Eq. (\ref{0}) will evolve to
\begin{eqnarray}
&&|\Phi^{\pm}_{1}\rangle_{3,2}\otimes|\Omega_{1}\rangle\otimes|\Omega_{2}\rangle\otimes|\Omega_{3}\rangle\otimes|\Omega_{4}\rangle\nonumber\\
&\rightarrow&\frac{1}{16}[(|L\rangle_{a1}|L\rangle_{b1}|L\rangle_{c1}\pm|R\rangle_{a1}|R\rangle_{b1}|R\rangle_{c1})\otimes(|L\rangle_{a2}|L\rangle_{b2}|L\rangle_{c2}
\pm|R\rangle_{a2}|R\rangle_{b2}|R\rangle_{c2})\nonumber\\
&\otimes&(|g_{L}\rangle_{1}-|g_{R}\rangle_{1})\otimes(|g_{L}\rangle_{2}-|g_{R}\rangle_{2})
\otimes(|g_{L}\rangle_{3}-|g_{R}\rangle_{3})\otimes(|g_{L}\rangle_{4}-|g_{R}\rangle_{4})\nonumber\\
&+&(|L\rangle_{a1}|R\rangle_{b1}|L\rangle_{c1}\pm|R\rangle_{a1}|L\rangle_{b1}|R\rangle_{c1})\otimes(|L\rangle_{a2}|R\rangle_{b2}|L\rangle_{c2}
\pm|R\rangle_{a2}|L\rangle_{b2}|R\rangle_{c2})\nonumber\\
&\otimes&(|g_{L}\rangle_{1}+|g_{R}\rangle_{1})\otimes(|g_{L}\rangle_{2}+|g_{R}\rangle_{2})
\otimes(|g_{L}\rangle_{3}+|g_{R}\rangle_{3})\otimes(|g_{L}\rangle_{4}+|g_{R}\rangle_{4})\nonumber\\
&+&(|L\rangle_{a1}|L\rangle_{b1}|R\rangle_{c1}\mp|R\rangle_{a1}|R\rangle_{b1}|L\rangle_{c1})\otimes(|L\rangle_{a2}|L\rangle_{b2}|R\rangle_{c2}
\mp|R\rangle_{a2}|R\rangle_{b2}|L\rangle_{c2})\nonumber\\
&\otimes&(|g_{L}\rangle_{1}-|g_{R}\rangle_{1})\otimes(|g_{L}\rangle_{2}-|g_{R}\rangle_{2})
\otimes(|g_{L}\rangle_{3}+|g_{R}\rangle_{3})\otimes(|g_{L}\rangle_{4}+|g_{R}\rangle_{4})\nonumber\\
&+&(|L\rangle_{a1}|R\rangle_{b1}|R\rangle_{c1}\mp|R\rangle_{a1}|L\rangle_{b1}|L\rangle_{c1})\otimes(|L\rangle_{a2}|R\rangle_{b2}|R\rangle_{c2}
\mp|R\rangle_{a2}|L\rangle_{b2}|L\rangle_{c2})\nonumber\\
&\otimes&(|g_{L}\rangle_{1}+|g_{R}\rangle_{1})\otimes(|g_{L}\rangle_{2}+|g_{R}\rangle_{2})
\otimes(|g_{L}\rangle_{3}-|g_{R}\rangle_{3})\otimes(|g_{L}\rangle_{4}-|g_{R}\rangle_{4})].\label{1}
\end{eqnarray}

Similarly, we can also obtain the other six cases. If the initial logic GHZ state is $|\Phi^{\pm}_{2}\rangle_{3,2}$, after the cavities, we can obtain
\begin{eqnarray}
&&|\Phi^{\pm}_{2}\rangle_{3,2}\otimes|\Omega_{1}\rangle\otimes|\Omega_{2}\rangle\otimes|\Omega_{3}\rangle\otimes|\Omega_{4}\rangle\nonumber\\
&\rightarrow&\frac{1}{16}[(|L\rangle_{a1}|L\rangle_{b1}|L\rangle_{c1}\pm|R\rangle_{a1}|R\rangle_{b1}|R\rangle_{c1})
\otimes(|R\rangle_{a2}|L\rangle_{b2}|L\rangle_{c2}\mp|L\rangle_{a2}|R\rangle_{b2}|R\rangle_{c2})\nonumber\\
&\otimes&(|g_{L}\rangle_{1}-|g_{R}\rangle_{1})\otimes(|g_{L}\rangle_{2}+|g_{R}\rangle_{2})
\otimes(|g_{L}\rangle_{3}-|g_{R}\rangle_{3})\otimes(|g_{L}\rangle_{4}-|g_{R}\rangle_{4})\nonumber\\
&+&(|R\rangle_{a1}|L\rangle_{b1}|L\rangle_{c1}\mp|L\rangle_{a1}|R\rangle_{b1}|R\rangle_{c1})
\otimes(|L\rangle_{a2}|L\rangle_{b2}|L\rangle_{c2}\pm|R\rangle_{a2}|R\rangle_{b2}|R\rangle_{c2})\nonumber\\
&\otimes&(|g_{L}\rangle_{1}+|g_{R}\rangle_{1})\otimes(|g_{L}\rangle_{2}-|g_{R}\rangle_{2})
\otimes(|g_{L}\rangle_{3}-|g_{R}\rangle_{3})\otimes(|g_{L}\rangle_{4}-|g_{R}\rangle_{4})\nonumber\\
&+&(|L\rangle_{a1}|R\rangle_{b1}|L\rangle_{c1}\pm|R\rangle_{a1}|L\rangle_{b1}|R\rangle_{c1})
\otimes(-|R\rangle_{a2}|R\rangle_{b2}|L\rangle_{c2}\pm|L\rangle_{a2}|L\rangle_{b2}|R\rangle_{c2})\nonumber\\
&\otimes&(|g_{L}\rangle_{1}+|g_{R}\rangle_{1})\otimes(|g_{L}\rangle_{2}-|g_{R}\rangle_{2})
\otimes(|g_{L}\rangle_{3}+|g_{R}\rangle_{3})\otimes(|g_{L}\rangle_{4}+|g_{R}\rangle_{4})\nonumber\\
&+&(|L\rangle_{a1}|L\rangle_{b1}|R\rangle_{c1}\mp|R\rangle_{a1}|R\rangle_{b1}|L\rangle_{c1})
\otimes(|R\rangle_{a2}|L\rangle_{b2}|R\rangle_{c2}\pm|L\rangle_{a2}|R\rangle_{b2}|L\rangle_{c2})\nonumber\\
&\otimes&(|g_{L}\rangle_{1}-|g_{R}\rangle_{1})\otimes(|g_{L}\rangle_{2}+|g_{R}\rangle_{2})
\otimes(|g_{L}\rangle_{3}+|g_{R}\rangle_{3})\otimes(|g_{L}\rangle_{4}+|g_{R}\rangle_{4})].\label{2}
\end{eqnarray}

For $|\Phi^{\pm}_{3}\rangle_{3,2}$, the photon state combined with four atom states will evolve to
\begin{eqnarray}
&&|\Phi^{\pm}_{3}\rangle_{3,2}\otimes|\Omega_{1}\rangle\otimes|\Omega_{2}\rangle\otimes|\Omega_{3}\rangle\otimes|\Omega_{4}\rangle\nonumber\\
&\rightarrow&\frac{1}{16}[(|L\rangle_{a1}|L\rangle_{b1}|L\rangle_{c1}\pm|R\rangle_{a1}|R\rangle_{b1}|R\rangle_{c1})
\otimes(|L\rangle_{a2}|R\rangle_{b2}|L\rangle_{c2}\mp|R\rangle_{a2}|L\rangle_{b2}|R\rangle_{c2})\nonumber\\
&\otimes&(|g_{L}\rangle_{1}-|g_{R}\rangle_{1})\otimes(|g_{L}\rangle_{2}+|g_{R}\rangle_{2})
\otimes(|g_{L}\rangle_{3}-|g_{R}\rangle_{3})\otimes(|g_{L}\rangle_{4}+|g_{R}\rangle_{4})\nonumber\\
&+&(|L\rangle_{a1}|R\rangle_{b1}|L\rangle_{c1}\pm|R\rangle_{a1}|R\rangle_{b1}|L\rangle_{c1})
\otimes(|L\rangle_{a2}|L\rangle_{b2}|L\rangle_{c2}\pm|R\rangle_{a2}|R\rangle_{b2}|R\rangle_{c2})\nonumber\\
&\otimes&(|g_{L}\rangle_{1}+|g_{R}\rangle_{1})\otimes(|g_{L}\rangle_{2}-|g_{R}\rangle_{2})
\otimes(|g_{L}\rangle_{3}+|g_{R}\rangle_{3})\otimes(|g_{L}\rangle_{4}-|g_{R}\rangle_{4})\nonumber\\
&+&(|L\rangle_{a1}|L\rangle_{b1}|R\rangle_{c1}\mp|R\rangle_{a1}|R\rangle_{b1}|L\rangle_{c1})
\otimes(-|L\rangle_{a2}|R\rangle_{b2}|R\rangle_{c2}\pm|R\rangle_{a2}|L\rangle_{b2}|L\rangle_{c2})\nonumber\\
&\otimes&(|g_{L}\rangle_{1}-|g_{R}\rangle_{1})\otimes(|g_{L}\rangle_{2}+|g_{R}\rangle_{2})
\otimes(|g_{L}\rangle_{3}+|g_{R}\rangle_{3})\otimes(|g_{L}\rangle_{4}-|g_{R}\rangle_{4})\nonumber\\
&+&(|L\rangle_{a1}|R\rangle_{b1}|R\rangle_{c1}\mp|R\rangle_{a1}|L\rangle_{b1}|L\rangle_{c1})
\otimes(-|L\rangle_{a2}|L\rangle_{b2}|R\rangle_{c2}\pm|R\rangle_{a2}|R\rangle_{b2}|L\rangle_{c2})\nonumber\\
&\otimes&(|g_{L}\rangle_{1}+|g_{R}\rangle_{1})\otimes(|g_{L}\rangle_{2}-|g_{R}\rangle_{2})
\otimes(|g_{L}\rangle_{3}-|g_{R}\rangle_{3})\otimes(|g_{L}\rangle_{4}+|g_{R}\rangle_{4})].\label{3}
\end{eqnarray}

Finally, $|\Phi^{\pm}_{4}\rangle_{3,2}$ combined with four atom states will evolve to
\begin{eqnarray}
&&|\Phi^{\pm}_{4}\rangle_{3,2}\otimes|\Omega_{1}\rangle\otimes|\Omega_{2}\rangle\otimes|\Omega_{3}\rangle\otimes|\Omega_{4}\rangle\nonumber\\
&\rightarrow&\frac{1}{16}[(|L\rangle_{a1}|L\rangle_{b1}|L\rangle_{c1}\pm|R\rangle_{a1}|R\rangle_{b1}|R\rangle_{c1})
\otimes(|L\rangle_{a2}|L\rangle_{b2}|R\rangle_{c2}\mp|R\rangle_{a2}|R\rangle_{b2}|L\rangle_{c2})\nonumber\\
&\otimes&(|g_{L}\rangle_{1}-|g_{R}\rangle_{1})\otimes(|g_{L}\rangle_{2}-|g_{R}\rangle_{2})
\otimes(|g_{L}\rangle_{3}-|g_{R}\rangle_{3})\otimes(|g_{L}\rangle_{4}+|g_{R}\rangle_{4})\nonumber\\
&+&(|L\rangle_{a1}|L\rangle_{b1}|R\rangle_{c1}\mp|R\rangle_{a1}|R\rangle_{b1}|L\rangle_{c1})
\otimes(|L\rangle_{a2}|L\rangle_{b2}|L\rangle_{c2}\pm|R\rangle_{a2}|R\rangle_{b2}|R\rangle_{c2})\nonumber\\
&\otimes&(|g_{L}\rangle_{1}-|g_{R}\rangle_{1})\otimes(|g_{L}\rangle_{2}-|g_{R}\rangle_{2})
\otimes(|g_{L}\rangle_{3}+|g_{R}\rangle_{3})\otimes(|g_{L}\rangle_{4}-|g_{R}\rangle_{4})\nonumber\\
&+&(|L\rangle_{a1}|R\rangle_{b1}|L\rangle_{c1}\pm|R\rangle_{a1}|L\rangle_{b1}|R\rangle_{c1})
\otimes(-|L\rangle_{a2}|R\rangle_{b2}|R\rangle_{c2}\pm|R\rangle_{a2}|L\rangle_{b2}|L\rangle_{c2})\nonumber\\
&\otimes&(|g_{L}\rangle_{1}+|g_{R}\rangle_{1})\otimes(|g_{L}\rangle_{2}+|g_{R}\rangle_{2})
\otimes(|g_{L}\rangle_{3}+|g_{R}\rangle_{3})\otimes(|g_{L}\rangle_{4}-|g_{R}\rangle_{4})\nonumber\\
&+&(-|L\rangle_{a1}|R\rangle_{b1}|R\rangle_{c1}\pm|R\rangle_{a1}|L\rangle_{b1}|L\rangle_{c1})
\otimes(|L\rangle_{a2}|R\rangle_{b2}|L\rangle_{c2}\pm|R\rangle_{a2}|L\rangle_{b2}|R\rangle_{c2})\nonumber\\
&\otimes&(|g_{L}\rangle_{1}+|g_{R}\rangle_{1})\otimes(|g_{L}\rangle_{2}+|g_{R}\rangle_{2})
\otimes(|g_{L}\rangle_{3}-|g_{R}\rangle_{3})\otimes(|g_{L}\rangle_{4}+|g_{R}\rangle_{4})].\label{4}
\end{eqnarray}

After all the photons exiting the cavities, we perform the Hadamard operations on the atoms "1", "2", "3", and "4", respectively. The Hadamard operation will make $|g_{L}\rangle\rightarrow\frac{1}{\sqrt{2}}(|g_{L}\rangle+|g_{R}\rangle)$, and  $|g_{R}\rangle\rightarrow\frac{1}{\sqrt{2}}(|g_{L}\rangle-|g_{R}\rangle)$. After that, we measure the atom states of the four atoms. From Eqs. (\ref{1})-(\ref{4}), it can be found that under the case that the initial photon state is $|\Phi^{\pm}_{1}\rangle_{3,2}$, the measurement results of the atoms "1" and "2", "3" and "4" are always the same, that is, if the measurement result of atom "1" ("3") is $|g_{L}\rangle$, the measurement result of atom "2" ("4") is also $|g_{L}\rangle$, while if the measurement result of atom "1" ("3") is $|g_{R}\rangle$, that of atom "2" ("4") must be $|g_{R}\rangle$. For $|\Phi^{\pm}_{2}\rangle_{3,2}$, the measurement results of atom "1" and "2" are different, that is, if the measurement result of atom "1" is $|g_{L}\rangle$, the measurement result of atom "2" must be $|g_{R}\rangle$, and vice versa. On the other hand, the measurement results of atom "3" and "4" are the same. For the initial state of $|\Phi^{\pm}_{3}\rangle_{3,2}$, both the measurement results of the atoms "1" and "2", "3" and "4" are different. For the initial state of $|\Phi^{\pm}_{4}\rangle_{3,2}$, the measurement results of atom "1" and "2" are the same, while those of atom "3" and "4" are different. Therefore, according to the measurement results of the four atoms, we can divide the eight C-GHZ states in Eq. (\ref{multi1}) into four groups $\{|\Phi^{\pm}_{1}\rangle_{3,2}\}$, $\{|\Phi^{\pm}_{2}\rangle_{3,2}\}$, $\{|\Phi^{\pm}_{3}\rangle_{3,2}\}$, and $\{|\Phi^{\pm}_{4}\rangle_{3,2}\}$. In detail, if the measurement result is $|g_{R}\rangle_{1}|g_{R}\rangle_{2}|g_{R}\rangle_{3}|g_{R}\rangle_{4}$, $|g_{L}\rangle_{1}|g_{L}\rangle_{2}|g_{L}\rangle_{3}|g_{L}\rangle_{4}$, $|g_{R}\rangle_{1}|g_{R}\rangle_{2}|g_{L}\rangle_{3}|g_{L}\rangle_{4}$, or $|g_{L}\rangle_{1}|g_{L}\rangle_{2}|g_{R}\rangle_{3}|g_{R}\rangle_{4}$, the initial state must be $|\Phi^{\pm}_{1}\rangle_{3,2}$. If the measurement result is  $|g_{R}\rangle_{1}|g_{L}\rangle_{2}|g_{R}\rangle_{3}|g_{R}\rangle_{4}$, $|g_{L}\rangle_{1}|g_{R}\rangle_{2}|g_{R}\rangle_{3}|g_{R}\rangle_{4}$, $|g_{L}\rangle_{1}|g_{R}\rangle_{2}|g_{L}\rangle_{3}|g_{L}\rangle_{4}$, or $|g_{R}\rangle_{1}|g_{L}\rangle_{2}|g_{L}\rangle_{3}|g_{L}\rangle_{4}$, the initial state must be $|\Phi^{\pm}_{2}\rangle_{3,2}$. On the other hand,
the measurement results $|g_{R}\rangle_{1}|g_{L}\rangle_{2}|g_{R}\rangle_{3}|g_{L}\rangle_{4}$, $|g_{L}\rangle_{1}|g_{R}\rangle_{2}|g_{L}\rangle_{3}|g_{R}\rangle_{4}$, $|g_{R}\rangle_{1}|g_{L}\rangle_{2}|g_{L}\rangle_{3}|g_{R}\rangle_{4}$, and $|g_{L}\rangle_{1}|g_{R}\rangle_{2}|g_{R}\rangle_{3}|g_{L}\rangle_{4}$ correspond to the initial state of $|\Phi^{\pm}_{3}\rangle_{3,2}$. The measurement results $|g_{R}\rangle_{1}|g_{R}\rangle_{2}|g_{R}\rangle_{3}|g_{L}\rangle_{4}$, $|g_{R}\rangle_{1}|g_{R}\rangle_{2}|g_{L}\rangle_{3}|g_{R}\rangle_{4}$, $|g_{L}\rangle_{1}|g_{L}\rangle_{2}|g_{R}\rangle_{3}|g_{L}\rangle_{4}$, and $|g_{L}\rangle_{1}|g_{L}\rangle_{2}|g_{L}\rangle_{3}|g_{R}\rangle_{4}$ correspond to the initial state of $|\Phi^{\pm}_{4}\rangle_{3,2}$.

In essence, the measurement results of the four atoms are determined by the parity characteristics of the input photon state. The four groups of C-GHZ states  $|\Phi^{\pm}_{1}\rangle_{3,2}$, $|\Phi^{\pm}_{2}\rangle_{3,2}$, $|\Phi^{\pm}_{3}\rangle_{3,2}$, and $|\Phi^{\pm}_{4}\rangle_{3,2}$ have different parity characteristics. For example, for $|\Phi^{\pm}_{1}\rangle_{3,2}$, the parities of the photons in $a_{1}b_{1}$ and $a_{2}b_{2}$ modes and the parities of the photons in $b_{1}c_{1}$ and $b_{2}c_{2}$ modes are both the same. That is, both even ($|L\rangle|L\rangle$ or $|R\rangle|R\rangle$) or both odd ($|L\rangle|R\rangle$ or $|R\rangle|L\rangle$). For $|\Phi^{\pm}_{2}\rangle_{3,2}$, the parities of the photons in $a_{1}b_{1}$ and $a_{2}b_{2}$ modes are different, while those of the photons in $b_{1}c_{1}$ and $b_{2}c_{2}$ modes are the same. For $|\Phi^{\pm}_{3}\rangle_{3,2}$, the parities of the photons in $a_{1}b_{1}$ and $a_{2}b_{2}$ modes and the parities of the photons in $b_{1}c_{1}$ and $b_{2}c_{2}$ modes are both different. For $|\Phi^{\pm}_{4}\rangle_{3,2}$, the parity of the photons in $a_{1}b_{1}$ and $a_{2}b_{2}$ modes are the same, while those of the photons in $b_{1}c_{1}$ and $b_{2}c_{2}$ modes are different. In this way, the first step of our protocol is essentially to make the parity check of the photon states with the help of the photonic Faraday rotation in low-Q cavity.

In the second step, we aim to distinguish the two C-GHZ states in each group. The discrimination processes of the four groups are quite similar. In this way, we take the discrimination of $|\Phi^{\pm}_{1}\rangle_{3,2}$ for example. After the atom states measurement in the first step, if the measurement result of the four atoms is $|g_{R}\rangle_{1}|g_{R}\rangle_{2}|g_{R}\rangle_{3}|g_{R}\rangle_{4}$, $|\Phi^{\pm}_{1}\rangle_{3,2}$ will collapse to
\begin{eqnarray}
|\Phi^{\pm}_{1}\rangle^{1}_{3,2}=\frac{1}{\sqrt{2}}(|L\rangle_{a1}|L\rangle_{b1}|L\rangle_{c1}\pm|R\rangle_{a1}|R\rangle_{b1}|R\rangle_{c1})
\otimes\frac{1}{\sqrt{2}}(|L\rangle_{a2}|L\rangle_{b2}|L\rangle_{c2}\pm|R\rangle_{a2}|R\rangle_{b2}|R\rangle_{c2}).\label{new1}
\end{eqnarray}
If the measurement result is $|g_{L}\rangle_{1}|g_{L}\rangle_{2}|g_{L}\rangle_{3}|g_{L}\rangle_{4}$, $|\Phi^{\pm}_{1}\rangle_{3,2}$ will collapse to
\begin{eqnarray}
|\Phi^{\pm}_{1}\rangle^{2}_{3,2}=\frac{1}{\sqrt{2}}(|L\rangle_{a1}|R\rangle_{b1}|L\rangle_{c1}\pm|R\rangle_{a1}|L\rangle_{b1}|R\rangle_{c1})
\otimes\frac{1}{\sqrt{2}}(|L\rangle_{a2}|R\rangle_{b2}|L\rangle_{c2}
\pm|R\rangle_{a2}|L\rangle_{b2}|R\rangle_{c2}).\label{new2}
\end{eqnarray}
For the measurement result of $|g_{R}\rangle_{1}|g_{R}\rangle_{2}|g_{L}\rangle_{3}|g_{L}\rangle_{4}$, $|\Phi^{\pm}_{1}\rangle_{3,2}$ will collapse to
\begin{eqnarray}
|\Phi^{\pm}_{1}\rangle^{3}_{3,2}=\frac{1}{\sqrt{2}}(|L\rangle_{a1}|L\rangle_{b1}|R\rangle_{c1}\mp|R\rangle_{a1}|R\rangle_{b1}|L\rangle_{c1})
\otimes\frac{1}{\sqrt{2}}(|L\rangle_{a2}|L\rangle_{b2}|R\rangle_{c2}\mp|R\rangle_{a2}|R\rangle_{b2}|L\rangle_{c2}).\label{new3}
\end{eqnarray}
Finally, for $|g_{L}\rangle_{1}|g_{L}\rangle_{2}|g_{R}\rangle_{3}|g_{R}\rangle_{4}$, $|\Phi^{\pm}_{1}\rangle_{3,2}$ will collapse to
\begin{eqnarray}
|\Phi^{\pm}_{1}\rangle^{4}_{3,2}=\frac{1}{\sqrt{2}}(|L\rangle_{a1}|R\rangle_{b1}|R\rangle_{c1}\mp|R\rangle_{a1}|L\rangle_{b1}|L\rangle_{c1})
\otimes\frac{1}{\sqrt{2}}(|L\rangle_{a2}|R\rangle_{b2}|R\rangle_{c2}\mp|R\rangle_{a2}|L\rangle_{b2}|L\rangle_{c2}).\label{new4}
\end{eqnarray}

Next, we take the operation on Eq. (\ref{new1}) for example. It can be found that we only need to distinguish $\frac{1}{\sqrt{2}}(|L\rangle_{a1}|L\rangle_{b1}|L\rangle_{c1}\pm|R\rangle_{a1}|R\rangle_{b1}|R\rangle_{c1})$ or $\frac{1}{\sqrt{2}}(|L\rangle_{a2}|L\rangle_{b2}|L\rangle_{c2}\pm|R\rangle_{a2}|R\rangle_{b2}|R\rangle_{c2})$. Here, we choose to distinguish $\frac{1}{\sqrt{2}}(|L\rangle_{a2}|L\rangle_{b2}|L\rangle_{c2}\pm|R\rangle_{a2}|R\rangle_{b2}|R\rangle_{c2})$. If we obtain other cases in Eqs. (\ref{new2})-(\ref{new4}), we can also transform the discrimination objects to $\frac{1}{\sqrt{2}}(|L\rangle_{a2}|L\rangle_{b2}|L\rangle_{c2}\pm|R\rangle_{a2}|R\rangle_{b2}|R\rangle_{c2})$ with the help of the bit-flip operation. For example, if we get $|\Phi^{\pm}_{1}\rangle^{2}_{3,2}$, we need to distinguish $\frac{1}{\sqrt{2}}(|L\rangle_{a2}|R\rangle_{b2}|L\rangle_{c2}
\pm|R\rangle_{a2}|L\rangle_{b2}|R\rangle_{c2})$, which can be transformed to  $\frac{1}{\sqrt{2}}(|L\rangle_{a2}|L\rangle_{b2}|L\rangle_{c2}\pm|R\rangle_{a2}|R\rangle_{b2}|R\rangle_{c2})$ with the bit-flip operation on the photon in $b_{2}$ mode. For $|\Phi^{\pm}_{1}\rangle^{3}_{3,2}$ and $|\Phi^{\pm}_{1}\rangle^{4}_{3,2}$, we can obtain the same results with the similar operation.

For distinguishing $\frac{1}{\sqrt{2}}(|L\rangle_{a2}|L\rangle_{b2}|L\rangle_{c2}+|R\rangle_{a2}|R\rangle_{b2}|R\rangle_{c2})$ from $\frac{1}{\sqrt{2}}(|L\rangle_{a2}|L\rangle_{b2}|L\rangle_{c2}-|R\rangle_{a2}|R\rangle_{b2}|R\rangle_{c2})$, we first make the photons in the $a_{2}b_{2}c_{2}$ modes pass through three HWPs, respectively. After the HWPs, $\frac{1}{\sqrt{2}}(|L\rangle_{a2}|L\rangle_{b2}|L\rangle_{c2}\pm|R\rangle_{a2}|R\rangle_{b2}|R\rangle_{c2})$ will evolve to
\begin{eqnarray}
&&\frac{1}{\sqrt{2}}(|L\rangle_{a2}|L\rangle_{b2}|L\rangle_{c2}+|R\rangle_{a2}|R\rangle_{b2}|R\rangle_{c2})\nonumber\\
&\rightarrow&\frac{1}{2}(|L\rangle_{a2}|L\rangle_{b2}|L\rangle_{c2}+|L\rangle_{a2}|R\rangle_{b2}|R\rangle_{c2}
+|R\rangle_{a2}|L\rangle_{b2}|R\rangle_{c2}+|R\rangle_{a2}|R\rangle_{b2}|L\rangle_{c2}),\\
&&\frac{1}{\sqrt{2}}(|L\rangle_{a2}|L\rangle_{b2}|L\rangle_{c2}-|R\rangle_{a2}|R\rangle_{b2}|R\rangle_{c2})\\
&\rightarrow&\frac{1}{2}(|L\rangle_{a2}|L\rangle_{b2}|R\rangle_{c2}+|L\rangle_{a2}|R\rangle_{b2}|L\rangle_{c2}
+|R\rangle_{a2}|L\rangle_{b2}|L\rangle_{c2}+|R\rangle_{a2}|R\rangle_{b2}|R\rangle_{c2}).
\end{eqnarray}

Next, the parties make another three-level atom trapped in a low-Q cavity, here named atom "5". The atom is prepared in the ground state of $|\Omega_{5}\rangle=|g_{L}\rangle$. Then, they make the photons in $a_{2}b_{2}c_{2}$ modes successively enter the cavity and interact with atom "5". After the interaction, we can obtain
\begin{eqnarray}
&&\frac{1}{\sqrt{2}}(|L\rangle_{a2}|L\rangle_{b2}|L\rangle_{c2}+|R\rangle_{a2}|R\rangle_{b2}|R\rangle_{c2})\otimes|g_{L}\rangle_{5}\nonumber\\
&\rightarrow&\frac{1}{2}(-|L\rangle_{a2}|L\rangle_{b2}|L\rangle_{c2}
+|L\rangle_{a2}|R\rangle_{b2}|R\rangle_{c2}
+|R\rangle_{a2}|L\rangle_{b2}|R\rangle_{c2}
-|R\rangle_{a2}|R\rangle_{b2}|L\rangle_{c2})|g_{L}\rangle_{5},\\
&&\frac{1}{\sqrt{2}}(|L\rangle_{a2}|L\rangle_{b2}|L\rangle_{c2}-|R\rangle_{a2}|R\rangle_{b2}|R\rangle_{c2})\otimes|g_{L}\rangle_{5}\nonumber\\
&\rightarrow&\frac{1}{2}(|L\rangle_{a2}|L\rangle_{b2}|R\rangle_{c2}+|L\rangle_{a2}|R\rangle_{b2}|L\rangle_{c2}
+|R\rangle_{a2}|L\rangle_{b2}|L\rangle_{c2}-|R\rangle_{a2}|R\rangle_{b2}|R\rangle_{c2})(i|g_{L}\rangle_{5}).
\end{eqnarray}
Finally, the parties measure the state of atom "5". If the measurement result is $|g_{L}\rangle$, the discrimination object is $\frac{1}{\sqrt{2}}(|L\rangle_{a2}|L\rangle_{b2}|L\rangle_{c2}+|R\rangle_{a2}|R\rangle_{b2}|R\rangle_{c2})$, while if the measurement result is $i|g_{L}\rangle$, the discrimination object is $\frac{1}{\sqrt{2}}(|L\rangle_{a2}|L\rangle_{b2}|L\rangle_{c2}-|R\rangle_{a2}|R\rangle_{b2}|R\rangle_{c2})$.

So far, we can completely distinguish $|\Phi^{\pm}_{1}\rangle_{3,2}$. In detail, under the case that the measurement result in the first step is $|g_{R}\rangle_{1}|g_{R}\rangle_{2}|g_{R}\rangle_{3}|g_{R}\rangle_{4}$ or $|g_{L}\rangle_{1}|g_{L}\rangle_{2}|g_{L}\rangle_{3}|g_{L}\rangle_{4}$, if the measurement result in the second step is $|g_{L}\rangle$, the initial photon state is $|\Phi^{+}_{1}\rangle_{3,2}$, while if the measurement result in the second step is $i|g_{L}\rangle$, the initial photon state is $|\Phi^{-}_{1}\rangle_{3,2}$. Under the case that the measurement result in the first step is $|g_{R}\rangle_{1}|g_{R}\rangle_{2}|g_{L}\rangle_{3}|g_{L}\rangle_{4}$ or $|g_{L}\rangle_{1}|g_{L}\rangle_{2}|g_{R}\rangle_{3}|g_{R}\rangle_{4}$, the measurement result $|g_{L}\rangle$ in the second step corresponds to $|\Phi^{-}_{1}\rangle_{3,2}$, while $i|g_{L}\rangle$ corresponds to $|\Phi^{+}_{1}\rangle_{3,2}$.

For other three groups $\{|\Phi^{\pm}_{2}\rangle_{3,2}\}$, $\{|\Phi^{\pm}_{3}\rangle_{3,2}\}$, and $\{|\Phi^{\pm}_{4}\rangle_{3,2}\}$, the discrimination task can also be transformed to distinguish  $\frac{1}{\sqrt{2}}(|L\rangle_{a2}|L\rangle_{b2}|L\rangle_{c2}\pm|R\rangle_{a2}|R\rangle_{b2}|R\rangle_{c2})$ with the help of the bit-flip operation. For example, under the case that the measurement result in the first step is $|g_{R}\rangle_{1}|g_{L}\rangle_{2}|g_{R}\rangle_{3}|g_{R}\rangle_{4}$, $|\Phi^{\pm}_{2}\rangle_{3,2}$ will collapse to $\frac{1}{\sqrt{2}}(|L\rangle_{a1}|L\rangle_{b1}|L\rangle_{c1}\pm|R\rangle_{a1}|R\rangle_{b1}|R\rangle_{c1})
\otimes\frac{1}{\sqrt{2}}(|R\rangle_{a2}|L\rangle_{b2}|L\rangle_{c2}\mp|L\rangle_{a2}|R\rangle_{b2}|R\rangle_{c2})$. For distinguishing $|\Phi^{+}_{2}\rangle_{3,2}$ from $|\Phi^{-}_{2}\rangle_{3,2}$, we only need to distinguish $\frac{1}{\sqrt{2}}(|R\rangle_{a2}|L\rangle_{b2}|L\rangle_{c2}\mp|L\rangle_{a2}|R\rangle_{b2}|R\rangle_{c2})$. After performing the bit-flip operation on the photon in $a_{2}$ mode, we can transform $\frac{1}{\sqrt{2}}(|R\rangle_{a2}|L\rangle_{b2}|L\rangle_{c2}\mp|L\rangle_{a2}|R\rangle_{b2}|R\rangle_{c2})$ to $\frac{1}{\sqrt{2}}(|L\rangle_{a2}|L\rangle_{b2}|L\rangle_{c2}\mp|R\rangle_{a2}|R\rangle_{b2}|R\rangle_{c2})$. If we obtain  $|g_{L}\rangle_{1}|g_{R}\rangle_{2}|g_{R}\rangle_{3}|g_{R}\rangle_{4}$, $|g_{R}\rangle_{1}|g_{L}\rangle_{2}|g_{L}\rangle_{3}|g_{L}\rangle_{4}$ or $|g_{L}\rangle_{1}|g_{R}\rangle_{2}|g_{L}\rangle_{3}|g_{L}\rangle_{4}$ in the first step, we can also transform the discrimination objects in the second step to $\frac{1}{\sqrt{2}}(|L\rangle_{a2}|L\rangle_{b2}|L\rangle_{c2}\pm|R\rangle_{a2}|R\rangle_{b2}|R\rangle_{c2})$. Therefore, combined with the measurement results of both two steps, we can completely distinguish the eight C-GHZ states.

In above discrimination process, we suppose each logic qubit is encoded in a Bell state. Actually, the logic qubit can
be encoded in a generalized GHZ state, that is, $m>2$. In this way, the eight C-GHZ states in Eq. (\ref{multi1}) can be written as
\begin{eqnarray}
|\Phi^{\pm}_{1}\rangle_{3,m}&=&\frac{1}{\sqrt{2}}(|GHZ_{m}^{+}\rangle_{A}|GHZ_{m}^{+}\rangle_{B}|GHZ_{m}^{+}\rangle_{C}
\pm|GHZ_{m}^{-}\rangle_{A}|GHZ_{m}^{-}\rangle_{B}|GHZ_{m}^{-}\rangle_{C}),\nonumber\\
|\Phi^{\pm}_{2}\rangle_{3,m}&=&\frac{1}{\sqrt{2}}(|GHZ_{m}^{-}\rangle_{A}|GHZ_{m}^{+}\rangle_{B}|GHZ_{m}^{+}\rangle_{C}
\pm|GHZ_{m}^{+}\rangle_{A}|GHZ_{m}^{-}\rangle_{B}|GHZ_{m}^{-}\rangle_{C}),\nonumber\\
|\Phi^{\pm}_{3}\rangle_{3,m}&=&\frac{1}{\sqrt{2}}(|GHZ_{m}^{+}\rangle_{A}|GHZ_{m}^{-}\rangle_{B}|GHZ_{m}^{+}\rangle_{C}
\pm|GHZ_{m}^{-}\rangle_{A}|GHZ_{m}^{+}\rangle_{B}|GHZ_{m}^{-}\rangle_{C}),\nonumber\\
|\Phi^{\pm}_{4}\rangle_{3,m}&=&\frac{1}{\sqrt{2}}(|GHZ_{m}^{+}\rangle_{A}|GHZ_{m}^{+}\rangle_{B}|GHZ_{m}^{-}\rangle_{C}
\pm|GHZ_{m}^{-}\rangle_{A}|GHZ_{m}^{-}\rangle_{B}|GHZ_{m}^{+}\rangle_{C}).\label{multim}
\end{eqnarray}
Here,
\begin{eqnarray}
|GHZ_{m}^{\pm}\rangle=\frac{1}{\sqrt{2}}(|L\rangle^{\otimes m}\pm|R\rangle^{\otimes m}).
\end{eqnarray}

Our analysis protocol can also completely distinguish the C-GHZ states in Eq. (\ref{multim}). Before the discrimination process described above, we require to perform $m-2$ Hadamard operations on $m-2$ photons in each logic qubit.  After the Hadamard operations, $|GHZ_{m}^{\pm}\rangle$ will evolve to
\begin{eqnarray}
|GHZ_{m}^{\pm}\rangle\rightarrow
\frac{1}{\sqrt{2}}[|L\rangle|L\rangle(\frac{1}{\sqrt{2}}(|L\rangle+|R\rangle))^{\otimes m-2}
\pm|R\rangle|R\rangle(\frac{1}{\sqrt{2}}(|L\rangle-|R\rangle))^{\otimes m-2}].
\end{eqnarray}
Then, we measure the $m-2$ photons in the basis of $\{|L\rangle, |R\rangle\}$.
If the number of $|R\rangle$ is even,
$|GHZ_{m}^{\pm}\rangle$ will collapse to $|\phi^{\pm}\rangle$, while if the number of $|R\rangle$ is odd, $|GHZ_{m}^{\pm}\rangle$ will become $|\phi^{\mp}\rangle$. In this way, we can finally transform the C-GHZ states in Eq. (\ref{multim}) to the C-GHZ states in Eq. (\ref{multi1}). It is noticed that in practical applications, we can increase the number of physical qubit $m$ to resist the environmental noise. This robust feature makes our analysis protocol quite useful in the point to point quantum communication based on logic-qubit entanglement.

\section{The complete C-GHZ state analysis with arbitrary $N$}
\begin{figure}[!h]
\begin{center}
\includegraphics[width=14cm,angle=0]{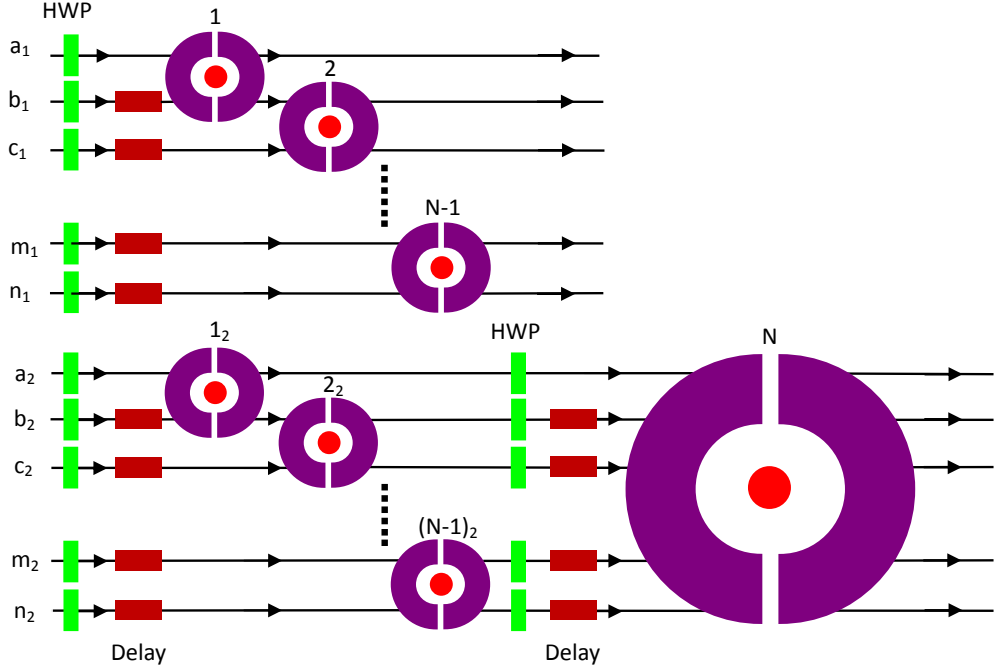}
\caption{The schematic drawing of our complete analysis protocol for the C-GHZ states with arbitrary $N$. The protocol also includes two steps. In the first step, the parties need to prepare $2(N-1)$ three-level atoms with the form of $\frac{1}{\sqrt{2}}(|g_{L}\rangle+|g_{R}\rangle)$ trapped in $2(N-1)$ low-Q cavities, respectively. In the second step, they need to prepare one three-level atom with the form of $|g_{L}\rangle$ trapped in another low-Q cavity. }
\end{center}
\end{figure}

Interestingly, our analysis protocol can be extended to completely distinguish the C-GHZ states with arbitrary $N$. For simplicity, we also suppose $m=2$. Under this case, there are $2^{N}$ C-GHZ states with the form of
\begin{eqnarray}
&&|\Phi^{\pm}_{1}\rangle_{N,2}=\frac{1}{\sqrt{2}}(|\phi^{+}\rangle^{\otimes N}\pm|\phi^{-}\rangle^{\otimes N}),\nonumber\\
&&|\Phi^{\pm}_{2}\rangle_{N,2}=\frac{1}{\sqrt{2}}(|\phi^{-}\rangle|\phi^{+}\rangle^{\otimes N-1}\pm|\phi^{+}\rangle|\phi^{-}\rangle^{\otimes N-1}),\nonumber\\
&&\cdots,\nonumber\\
&&|\Phi^{\pm}_{2^{N-1}}\rangle_{N,2}=\frac{1}{\sqrt{2}}(|\phi^{+}\rangle^{\otimes N-1}|\phi^{-}\rangle\pm|\phi^{-}\rangle^{\otimes N-1}|\phi^{+}\rangle).\label{multi2}
\end{eqnarray}

 The basic principle of the complete analysis protocol is shown in Fig. 3. The analysis protocol also includes two steps. In the first step, we first make each of the photons pass through a HWP, which transforms  $|\phi^{-}\rangle$ to $|\psi^{+}\rangle$, but keeps $|\phi^{+}\rangle$ constant. After the HWPs, the $2^{N}$ C-GHZ states in Eq. (\ref{multi2}) will evolve to
\begin{eqnarray}
&&|\Phi^{\pm}_{1}\rangle_{N,2}\rightarrow\frac{1}{\sqrt{2}}(|\phi^{+}\rangle^{\otimes N}\pm|\psi^{+}\rangle^{\otimes N}),\nonumber\\
&&|\Phi^{\pm}_{2}\rangle_{N,2}\rightarrow\frac{1}{\sqrt{2}}(|\psi^{+}\rangle|\phi^{+}\rangle^{\otimes N-1}\pm|\phi^{+}\rangle|\psi^{+}\rangle^{\otimes N-1}),\nonumber\\
&&\cdots,\nonumber\\
&&|\Phi^{\pm}_{2^{N-1}}\rangle_{N,2}\rightarrow\frac{1}{\sqrt{2}}(|\phi^{+}\rangle^{\otimes N-1}|\psi^{+}\rangle\pm|\psi^{+}\rangle^{\otimes N-1}|\phi^{+}\rangle).\label{multi3}
\end{eqnarray}

Then, the parties prepare $2(N-1)$ three-level atoms, named "$1$", "$2$", $\cdots$, "$N-1$", and "$1_{2}$", "$2_{2}$", $\cdots$, "$(N-1)_{2}$" trapped in $2(N-1)$ low-Q cavities, respectively. All the $2(N-1)$ atoms are prepared in the same state of $\frac{1}{\sqrt{2}}(|g_{L}\rangle+|g_{R}\rangle)$. The parties make the photons in the $a_{1}b_{1}$, $b_{1}c_{1}$, $\cdots$, $m_{1}n_{1}$ modes pass through $N-1$ cavities and interact with the atoms "$1$", "$2$", $\cdots$, "$N-1$", respectively. Meanwhile, they make the photons in the $a_{2}b_{2}$, $b_{2}c_{2}$, $\cdots$, $m_{2}n_{2}$ modes pass through another $N-1$ cavities and interact with the atoms "$1_{2}$", "$2_{2}$", $\cdots$, "$(N-1)_{2}$", respectively. The parties also adopt "Delay" setup to control the time of the photon entering the cavity for ensuring only a photon interact with the atom at a time. After all the photons exiting the cavities, we make the Hadamard operations on all the $2(N-1)$ atoms and measure them in the basis of $\{|g_{L}\rangle, |g_{R}\rangle\}$. Based on the parity features of the input photon states, the different initial states will cause different measurement results. According to the measurement results, we can divide the $2^{N}$ C-GHZ states in Eq. (\ref{multi3}) into $2^{N-1}$ groups, say $\{|\Phi^{\pm}_{1}\rangle_{N,2}\}$, $\{|\Phi^{\pm}_{2}\rangle_{N,2}\}$, $\cdots$, $\{|\Phi^{\pm}_{2^{N-1}}\rangle_{N,2}\}$. For example, for $\{|\Phi^{\pm}_{1}\rangle_{N,2}\}$, the measurement results of the atoms "$1$", "$2$", $\cdots$, "$N-1$" equal to those of the atoms "$1_{2}$", "$2_{2}$", $\cdots$, "$(N-1)_{2}$", respectively. For $\{|\Phi^{\pm}_{2}\rangle_{N,2}\}$, the measurement result of the atom "$1$" is different from that of the atom "$1_{2}$", while the measurement results of the atoms "$2$", $\cdots$, "$N-1$" equal to those of the atoms "$2_{2}$", $\cdots$, "$(N-1)_{2}$", respectively. For $\{|\Phi^{\pm}_{3}\rangle_{N,2}\}$, the measurement results of the atom "$1$" and "2" are different from those of the atom "$1_{2}$" and "$2_{2}$", respectively, while the measurement results of the atoms "$3$", $\cdots$, "$N-1$" equal to those of the atoms "$3_{2}$", $\cdots$, "$(N-1)_{2}$", respectively. $\cdots$

The second step is to distinguish the two states in each of the $2^{N-1}$ groups. We also take the discrimination of $|\Phi^{\pm}_{1}\rangle_{N,2}$ for example. For the initial states of $|\Phi^{\pm}_{1}\rangle_{N,2}$, we can obtain $2^{N-1}$ different measurement results in the first step, which can be written as $|g_{1}\rangle_{1}|g_{1}\rangle_{1_{2}}|g_{2}\rangle_{2}|g_{2}\rangle_{2_{2}}\cdots|g_{n-1}\rangle_{N-1}|g_{n-1}\rangle_{(N-1)_{2}}$, where $|g_{i}\rangle=|g_{L}\rangle$ or $|g_{R}\rangle$ $(i=1, 2, \cdots, n-1)$. When the measurement result is $|g_{R}\rangle_{1}|g_{R}\rangle_{1_{2}}|g_{R}\rangle_{2}|g_{R}\rangle_{2_{2}}|g_{R}\rangle_{3}|g_{R}\rangle_{3_{2}}\cdots|g_{R}\rangle_{N-1}|g_{R}\rangle_{(N-1)_{2}}$, the initial state will collapse to
\begin{eqnarray}
|\Phi^{\pm}_{1}\rangle_{N,2}&\rightarrow&\frac{1}{\sqrt{2}}(|L\rangle_{a_{1}}|L\rangle_{b_{1}}|L\rangle_{c_{1}}\cdots|L\rangle_{n_{1}}\pm
|R\rangle_{a_{1}}|R\rangle_{b_{1}}|R\rangle_{c_{1}}\cdots|R\rangle_{n_{1}})\nonumber\\
&\otimes&\frac{1}{\sqrt{2}}(|L\rangle_{a_{2}}|L\rangle_{b_{2}}|L\rangle_{c_{2}}
\cdots|L\rangle_{n_{2}}\pm
|R\rangle_{a_{2}}|R\rangle_{b_{2}}|R\rangle_{c_{2}}\cdots|R\rangle_{n_{2}}).
\end{eqnarray}
In this way, we only need to distinguish $\frac{1}{\sqrt{2}}(|L\rangle_{a_{2}}|L\rangle_{b_{2}}|L\rangle_{c_{2}}
\cdots|L\rangle_{n_{2}}\pm
|R\rangle_{a_{2}}|R\rangle_{b_{2}}|R\rangle_{c_{2}}\cdots|R\rangle_{n_{2}})$. For achieving the aim, the parties make the photons in the $a_{2}, b_{2}, c_{2}, \cdots, m_{2}, n_{2}$ modes pass through $N$ HWPs, respectively. After the HWPs, $\frac{1}{\sqrt{2}}(|L\rangle_{a_{2}}|L\rangle_{b_{2}}|L\rangle_{c_{2}}
\cdots|L\rangle_{n_{2}}\pm
|R\rangle_{a_{2}}|R\rangle_{b_{2}}|R\rangle_{c_{2}}\cdots|R\rangle_{n_{2}})$ will evolve to
 \begin{eqnarray}
 && \frac{1}{\sqrt{2}}(|L\rangle_{a_{2}}|L\rangle_{b_{2}}\cdots|L\rangle_{n_{2}}+|R\rangle_{a_{2}}|R\rangle_{b_{2}}\cdots|R\rangle_{n_{2}})
  \rightarrow(\frac{1}{\sqrt{2}})^{N+1}[(|L\rangle_{a_{2}}+|R\rangle_{a_{2}})(|L\rangle_{b_{2}}+|R\rangle_{b_{2}})
\cdots(|L\rangle_{n_{2}}+|R\rangle_{n_{2}})\nonumber\\
  &&+(|L\rangle_{a_{2}}-|R\rangle_{a_{2}})(|L\rangle_{b_{2}}-|R\rangle_{b_{2}})\cdots(|L\rangle_{n_{2}}-|R\rangle_{n_{2}})],
  \end{eqnarray}
  and
    \begin{eqnarray}
 && \frac{1}{\sqrt{2}}(|L\rangle_{a_{2}}|L\rangle_{b_{2}}\cdots|L\rangle_{n_{2}}-|R\rangle_{a_{2}}|R\rangle_{b_{2}}\cdots|R\rangle_{n_{2}})
  \rightarrow(\frac{1}{\sqrt{2}})^{N+1}[(|L\rangle_{a_{2}}+|R\rangle_{a_{2}})(|L\rangle_{b_{2}}+|R\rangle_{b_{2}})
  \cdots(|L\rangle_{n_{2}}+|R\rangle_{n_{2}})\nonumber\\
  &&-(|L\rangle_{a_{2}}-|R\rangle_{a_{2}})(|L\rangle_{b_{2}}-|R\rangle_{b_{2}})\cdots(|L\rangle_{n_{2}}-|R\rangle_{n_{2}})].
  \end{eqnarray}

Then, the parties prepare an atom named atom "N" in the ground state of $|\Omega_{N}\rangle=|g_{L}\rangle$ and make it trap in a low-Q cavity. The parties make all the photons after the HWPs successively pass through the low-Q cavity and interact with the atom "N". After the interaction, the parties measure the atom "N". If the atom state is $|g_{L}\rangle$, the photon state must be $\frac{1}{\sqrt{2}}(|L\rangle_{a_{2}}|L\rangle_{b_{2}}\cdots|L\rangle_{n_{2}}+|R\rangle_{a_{2}}|R\rangle_{b_{2}}\cdots|R\rangle_{n_{2}})$, and the initial photon state is $|\Phi^{+}_{1}\rangle_{N,2}$. If the atom state is $i|g_{L}\rangle$, the photon state is $\frac{1}{\sqrt{2}}(|L\rangle_{a_{2}}|L\rangle_{b_{2}}\cdots|L\rangle_{n_{2}}-|R\rangle_{a_{2}}|R\rangle_{b_{2}}\cdots|R\rangle_{n_{2}})$, and the initial photon state must be $|\Phi^{-}_{1}\rangle_{N,2}$. On the other hand, for $|\Phi^{\pm}_{1}\rangle_{N,2}$, if we obtain other measurement results in the first step, such as $|g_{L}\rangle_{1}|g_{L}\rangle_{1_{2}}|g_{R}\rangle_{2}|g_{R}\rangle_{2_{2}}\cdots|g_{R}\rangle_{N-1}|g_{R}\rangle_{(N-1)_{2}}$, the initial state will collapse to $\frac{1}{\sqrt{2}}(|R\rangle_{a1}|L\rangle_{b1}|L\rangle_{c1}\cdots|L\rangle_{n_{1}}\mp|L\rangle_{a1}|R\rangle_{b1}|R\rangle_{c1}\cdots|R\rangle_{n_{1}})
\otimes\frac{1}{\sqrt{2}}(|R\rangle_{a2}|L\rangle_{b2}|L\rangle_{c2}\cdots|L\rangle_{n_{2}}\mp|L\rangle_{a2}|R\rangle_{b2}|R\rangle_{c2}\cdots|R\rangle_{n_{2}})$. We only need to distinguish $\frac{1}{\sqrt{2}}(|R\rangle_{a2}|L\rangle_{b2}|L\rangle_{c2}\cdots|L\rangle_{n_{2}}\mp|L\rangle_{a2}|R\rangle_{b2}|R\rangle_{c2}\cdots|R\rangle_{n_{2}})$. In this way, after performing a bit-flip operation on the photon in the $a_{2}$ mode, we can transform $\frac{1}{\sqrt{2}}(|R\rangle_{a2}|L\rangle_{b2}|L\rangle_{c2}\cdots|L\rangle_{n_{2}}\mp|L\rangle_{a2}|R\rangle_{b2}|R\rangle_{c2}\cdots|R\rangle_{n_{2}})$  to$\frac{1}{\sqrt{2}}(|L\rangle_{a2}|L\rangle_{b2}|L\rangle_{c2}\cdots|L\rangle_{n_{2}}\mp|R\rangle_{a2}|R\rangle_{b2}|R\rangle_{c2}\cdots|R\rangle_{n_{2}})$. Therefor, for all the cases corresponding to $2^{N-1}$ measurement results, $|\Phi^{\pm}_{1}\rangle_{N,2}$ can be distinguished with the same approach.

If the initial state is $|\Phi^{\pm}_{k}\rangle_{N,2}$ $(k\neq1)$, in the second step, the parties can also transform the photon states which need to be measured into $\frac{1}{\sqrt{2}}(|L\rangle_{a2}|L\rangle_{b2}|L\rangle_{c2}\cdots|L\rangle_{n_{2}}\pm|R\rangle_{a2}|R\rangle_{b2}|R\rangle_{c2}\cdots|R\rangle_{n_{2}})$ with the help of the bit-flip operation. Therefore, all the $2^{N}$ C-GHZ states can be completely distinguished based on the measurement results of both two steps.

Certainly, our analysis protocol can also completely distinguish the C-GHZ states with arbitrary $N$ and $m$. As described above, before the first analysis step, the parties need to first perform $m-2$ Hadamard operation on the $m-2$ photons in each logic qubit and then measure the $m-2$ photons in the basis of $\{|L\rangle, |R\rangle\}$. After the measurement, $|GHZ_{m}^{\pm}\rangle$ will become $|\phi^{\pm}\rangle$ under the case that the number of $|R\rangle$ is even. $|GHZ_{m}^{\pm}\rangle$ will become $|\phi^{\mp}\rangle$ when the number of $|R\rangle$ is odd. In this way, we can transform $|\Phi^{\pm}_{k}\rangle_{N,m}$ into $|\Phi^{\pm}_{k}\rangle_{N,2}$, and realize the complete analysis with the principle of the section.

\section{Discussion and conclusion}

In the paper, we proposed a complete analysis protocol for the C-GHZ states with arbitrary $N$ and $m$. Our protocol includes two steps. Before the first step, we need to perform $m-2$ Hadamard operation on $m-2$ photons in each logic qubit, and measure them in the basis of $\{|L\rangle, |R\rangle\}$. After the measurement, we can transform the C-GHZ states with arbitrary $m$ $(m>2)$ into the C-GHZ states with $m=2$. In both two steps, we essentially make parity check for the photon states with the help of the photo-atom interaction in low-Q cavities. In the first step, we require $2(N-1)$ three-level atoms trapped in $2(N-1)$ low-Q cavities, respectively, and make all the photons in the $2N$ spatial modes enter the cavities to interact with the atoms, respectively. After the photon-atom interaction, we can divide the $2^{N}$ C-GHZ states into $2^{N-1}$ groups according to measurement results of the atomic states. In the second step, we only require one three-level atom trapped in a low-Q cavity, and make the $N$ photons in the $a_{2}b_{2}c_{2}\cdots n_{2}$ modes enter the cavity and interact with the atom, successively. After the photon-atom interaction, we measure the atom state. In this way, we can distinguish the two states in each of the $2^{N-1}$ groups based on the measurement results of the atomic state. Therefore, we can completely distinguish the $2^{N}$ C-GHZ states according to the atomic states measurement results in both two analysis steps.

In our protocol, we essentially construct the parity check gate based on the photonic Faraday rotation in low-Q cavity. Fortunately, the photonic Faraday rotation in low-Q cavity has been experimentally realized in some recent research works. For example, in 2005, the group of Nu$\beta$mann once reported their experiment to precisely control and adjust
the individual ultracold $^{85}Rb$ atoms coupled to a high-finesse optical cavity \cite{Rb1}. The states of $|F\rangle=2$. $m_{F}=\pm1$ of the $5S_{1/2}$ are chosen to be the two ground
states $|g_{L}\rangle$ and $|g_{R}\rangle$, respectively. The transition frequency between the
ground states and the excited state at $\lambda=780 nm$ is controlled as $\omega_{0}=\frac{2 \pi c}{\lambda} \approx 2.42 \times 10^{15} Hz$, and the
cavity length, cavity rate and the finesse are kept as $L=38.6 \mu m$, $K = 2\pi \times 53 MHz$ and $F=37000$,
respectively. Later, Fortier \emph{et al.} experimentally realized the deterministic loading of single $^{87}Rb$ atoms into the cavity by incorporating a deterministic loaded atom conveyor \cite{Rb2}. In the same year, Colombe \emph{et al.} also reported their experiment on realizing the strong atom-field coupling for Bose-Einstein condensates (BEC) in a fiber-based
Fabry-Perot (F-P) cavity on a chip \cite{Rb3}. They showed that the $^{87}Rb$ BEC can be positioned deterministically anywhere within
the cavity and localized entirely within a single antinode of the standing-wave cavity field. Based on their attractive experimental results, our analysis protocol may be realized experimentally in the near future.

In theory, our analysis protocol can completely distinguish the $2^{N}$ C-GHZ states with the success probability of $100\%$. However, in practical experimental conditions, some imperfections are still existed. For example, the first kind of imperfection may come from the practical cavity condition. In our protocol, in order to obtain the input-output relationship as Eq. (\ref{relation}), the frequency of the input coherent state $\omega_{p}$ should satisfy $\omega_{p}=\omega_{c}-\frac{\kappa}{2}=\omega_{0}-\frac{\kappa}{2}$, and the photon-atom coupling strength should meet $\lambda=\frac{\kappa}{2}$. In practical experimental condition, the deviation of resonance $\triangle_{1}=\omega_{c}-\omega_{0}$,
and mismatch of coupling strength $\triangle_{2}=g-\kappa/2$ may change the phase shift of photonic Faraday rotation as $\phi-\phi_{0}=\pi/2+\sigma$, where $\sigma$ is a small quantity. Based on the calculation of Ref. \cite{zhoupra2}, the exist of $\sigma$ will cause errors in both two analysis steps. Fortunately, based on the experimental results from Refs. \cite{Rb1,Rb2,Rb3}, we can control the reflectivity of the input coherent state by manipulating the position of a single $^{87}Rb$ and precisely tuning the atom-cavity coupling strength. In this way, the value of $\sigma$ can be controlled very small, and the error probability ($P_{error}$) can be controlled quite low. On the other hand, the imperfections may also come from the detection efficiency for both the single-photon and single-atom measurement. The detection
efficiency indicates when the single photon (atom) enters the single-photon (atom) detector, the detector can register it
with the probability of $\eta$. In current experimental condition, neither the single-photon detection efficiency ($\eta_{p}$) nor the single-atom
detection efficiency ($\eta_{a}$) can reach 100\%.
 In our protocol, we need to measure the states of $m-2$ photons in each logic qubit prior to the discrimination and measure the states of $2N-1$ three-level atoms in two steps. If the photon (atom) detector does not register the photon (atom), it will cause a failure of our protocol. According to the above two imperfection factors, we can calculate the total success probability ($P_{N,m}$) of our protocol as
\begin{eqnarray}
P_{N,m}=(\eta_{p})^{N(m-2)}(\eta_{a})^{2N-1}(1-P_{error}).
 \end{eqnarray}
Recently, many research works about the single-atom and single-photon detections have been reported. For example, in 2010, Heine \emph{et al.} reported that they have achieved $\eta_{a}=66\%$ in the experiment. Moreover, they have shown that with some improvement, the single atom detection efficiency can achieve  $\eta_{a}>95\%$ in theory \cite{atomefficiency}. The single photon detection has been a challenge under current experimental
conditions, for the quantum decoherence effect of the photon detector \cite{photonefficiency}. Lita \emph{et al.} reported their experimental
result about the near-infrared single-photon detection. They showed the $\eta_{p}$ at 1556 $nm$ can reach $95\% \pm 2\%$ \cite{photonefficiency1}. Based on their results, we make the numerical simulation on the $P_{N,m}$ of our protocol. We assume $\eta_{a}=\eta_{p}=90\%$ and $P_{error}=0.05$. In this way, we can calculate that $P_{3,2}\approx0.561$, $P_{3,3}\approx0.409$, and $P_{4,3}\approx0.298$. Therefore, it can be found that with the growth of $N$ and $m$, $P_{N,m}$ will decrease largely.

In conclusion, we proposed a completely analysis protocol for the arbitrary C-GHZ states with the help of some auxiliary single atoms trapped in the low-Q cavities. The protocol includes two steps. In both two steps, we essentially adopt the QED in low-Q cavity, say the photonic Faraday rotation to perform the parity check for the photonic states. In the first step, we can divide the $2^{N}$ C-GHZ states into $2^{N-1}$ groups according to different measurement results of the $2(N-1)$ auxiliary single atoms. In the second step, we can distinguish the two C-GHZ states in each of $2^{N-1}$ groups based on the measurement result of another auxiliary single atom. In this way, our protocol can completely distinguish all the $2^{N}$ C-GHZ states. In ideal experimental condition, the success probability of our protocol can reach 100\%, and the number of
physical qubit encoded in each logic qubit does not affect the protocol. As we can increase the number of physical qubit in each
logic qubit to resist the environmental noise, this robust
feature make our protocol quite useful in the point to point quantum communication
based on logic-qubit entanglement.

\section*{ACKNOWLEDGEMENTS} This work was supported by the National Natural Science Foundation
of China under Grant  Nos. 11474168 and 61401222, the Natural Science Foundation of Jiangsu province under Grant No. BK20151502
, the Qing Lan Project in Jiangsu Province, and A Project
Funded by the Priority Academic Program Development of Jiangsu
Higher Education Institutions.


\begin{thebibliography}{0}
\bibitem{t1} Bennett C H, Brassard G, Crepeau C, Jozsa R, Peres A and Wootters W K 1993 \emph{Phys. Rev. Lett.} \textbf{70} 1895


\bibitem{QKD1}  Ekert A K 1991 \emph{Phys. Rev. Lett.} \textbf{67} 661

\bibitem{QSDC1}Long G L and Liu X S 2002 \emph{Phys. Rev. A}  \textbf{65} 032302
\bibitem{QSDC2}Deng F G,  Long G L and Liu X S 2003 \emph{Phys. Rev. A } \textbf{68} 042317

\bibitem{repeater} Briegel H J, D$\ddot{u}$r W, Cirac J I and Zoller P 1998 \emph{Phys. Rev. Lett.} \textbf{81} 5932

\bibitem{GHZ1} Sagi Y 2003 \emph{Phys. Rev. A} \textbf{68} 042320
\bibitem{GHZ2}  Deng Z J, Feng M and Gao K L 2007 \emph{Phys. Rev. A } \textbf{75} 024302
\bibitem{GHZ3} Yu C S, Yi X X, Song H S and Mei D 2007 \emph{Phys. Rev. A} \textbf{75} 044301
\bibitem{GHZ4} Xia Y, Lu P M and Zeng Y Z 2012 \emph{Quantum Infor. Process.} \textbf{11} 605

\bibitem{EPP1}  Pan J W, Simon C and  Zellinger A 2001 \emph{Nature} \textbf{410} 1067
\bibitem{EPP2}  Sheng Y B and  Deng F G 2010 \emph{Phys. Rev. A} \textbf{81} 032307
\bibitem{EPP3}  Gonta D and van Loock P 2011 \emph{Phys. Rev. A} \textbf{84} 042303
\bibitem{EPP4}  Sheng Y B, Zhou L and  Long G L 2013 \emph{Phys. Rev. A} \textbf{88} 022302
\bibitem{EPP5}  Wang C, Zhang Y and Jin G S 2011 \emph{Phys. Rev. A} \textbf{84} 032307
\bibitem{EPP6}  Zwerger M, Briegel H J and D$\ddot{u}$r W 2013 \emph{Phys.Rev. Lett.} \textbf{110} 260503
\bibitem{EPP7}  Zwerger M,  Briegel H J and D$\ddot{u}$r W 2014 \emph{Phys. Rev. A} \textbf{90} 012314
\bibitem{EPP8}  Ren B C, Du F F and Deng F G 2014 \emph{Phys. Rev. A }\textbf{90} 052309
\bibitem{EPP9}  Sheng Y B and  Zhou L 2015 \emph{Sci. Rep.} \textbf{5} 7815
\bibitem{EPP10} Sheng Y B and  Zhou L 2014 \emph{Laser Phys. Lett.} \textbf{11} 085203

\bibitem{ECP1} Yamamoto T, Koashi M and Imoto N 2001 \emph{Phys. Rev. A} \textbf{64} 012304
\bibitem{ECP2} Zhao Z, Pan J W and Zhan M S 2001 \emph{Phys. Rev. A} \textbf{64} 014301
\bibitem{ECP3}  Sheng Y B, Zhou L, Zhao S M and Zheng B Y 2012 \emph{Phys. Rev. A} \textbf{85} 012307
\bibitem{ECP4}  Deng F G 2012 \emph{Phys. Rev. A } \textbf{85} 022311
\bibitem{ECP5}  Zhou L,  Sheng Y B,  Cheng W W,  Gong L Y and  Zhao S M 2013 \emph{J. Opt. Soc. Am. B }\textbf{ 30} 71
\bibitem{ECP6} Wang C 2012 \emph{Phys. Rev. A} \textbf{86} 012323
\bibitem{ECP7} Zhou L, Sheng Y B and Wang X F 2013 \emph{J. Opt. Soc. Am. B} \textbf{31} 503

\bibitem{zhouNLA} Zhou L and  Sheng Y B 2015 \emph{Laser Phys. Lett.} \textbf{12} 045203
\bibitem{zhoujosa2} Zhou L and  Sheng Y B 2013 \emph{J. Opt. Soc. Am. B } \textbf{30} 2737

\bibitem{amplification} Xiang G Y, Ralph T C, Walk N and  Pryde G J 2010
\emph{Nat. Photon.} \textbf{4} 316

\bibitem{logic} Fr$\ddot{o}$wis F and D$\ddot{u}$r W 2011 \emph{Phys. Rev. Lett.} \textbf{106} 110402
 \bibitem{cghz2} Fr\"{o}wis F and D\"{u}r W 2012 \emph{Phys. Rev. A} \textbf{85} 052329
\bibitem{logic1} Lu H, Chen L K, Liu C,  Yan X C, Li L,  Liu N L, Zhao B, Chen Y A and  Pan J W 2014  \emph{Nat. Photon.} \textbf{8} 364
\bibitem{logic2}Ding D, Yan F L and Gao T 2013 \emph{J. Opt. Soc. Am. B} \textbf{30} 3075
\bibitem{logic3} Kesting F, Fr$\ddot{o}$wis F and D$\ddot{u}$r W 2013 \emph{Phys. Rev. A} \textbf{88} 042305
\bibitem{logic4} D$\ddot{u}$r W, Skotinioti M, Fr$\ddot{o}$wis F and Kraus B 2014 \emph{Phys. Rev. Lett.} \textbf{112} 080801



\bibitem{BSA1} Vaidman L and Yoran N 1999 \emph{Phys. Rev. A} \textbf{59} 116
\bibitem{BSA2} L\"{u}tkenhaus N, Calsamiglia J and Suominen K A 1999 \emph{Phys. Rev. A} \textbf{59} 3295


\bibitem{hyper1}  Walborn S P,  P\'{a}dua  S and  Monken C H 2003 \emph{Phys. Rev. A} \textbf{68} 042313
\bibitem{hyper2} Schuck C, Huber G, Kurtsiefer C and Weinfurter H 2006 \emph{Phys. Rev. Lett.} \textbf{96} 190501
\bibitem{hyper3} Barbieri  M, Vallone G, Mataloni P and De Martini F 2007 \emph{Phys. Rev. A} \textbf{75} 042317
\bibitem{hyper4} Barreiro J T, Wei T C and Kwiat P G 2008 \emph{Nat. Phys.}  \textbf{4} 282

\bibitem{kerr1}  Barrett S D, Kok P, Nemoto K, Beausoleil R G, Munro W J and Spiller T P 2005 \emph{Phys. Rev. A} \textbf{71} 060302
\bibitem{kerr2}  Sheng Y B, Deng F G and  Long G L 2010 \emph{Phys. Rev. A} \textbf{82} 032318

\bibitem{dot1}  Ren B C, Wei H R, Hua M, Li T and  Deng F G 2012 \emph{Opt. Exp.} \textbf{20} 20664
\bibitem{dot2}  Wang T J, Lu Y and  Long G L 2012 \emph{Phys. Rev. A} \textbf{86} 042337

\bibitem{QND1}  Nemoto K and Munro W J 2004 \emph{Phys. Rev. Lett.}  \textbf{93} 250502


\bibitem{GHZA1}  Pan J W and Zeilinger A 1998 \emph{Phys. Rev. A} \textbf{57} 2208
\bibitem{GHZA2} L$\ddot{u}$tkenhaus N, Calsamiglia J and Suominen K A 1999 \emph{Phys. Rev. A} \textbf{59} 3295
\bibitem{GHZA3} Calsamiglia J 2002 \emph{Phys. Rev. A} \textbf{65} 030301(R)
\bibitem{GHZA4} Schuck C, Huber G, Kurtsiefer C and Weinfurter H 2006 \emph{Phys. Rev. Lett.} \textbf{96} 190501
\bibitem{GHZA5} van Houwelingen J A W, Brunner N, Beveratos A, Zbinden H and Gisin N 2006 \emph{Phys. Rev. Lett.} \textbf{96} 130502
\bibitem{GHZA7} Sheng Y B and Deng F G 2010 \emph{Phys. Rev. A} \textbf{81} 032307

\bibitem{shengepl} Sheng Y B and  Zhou L 2015 \emph{EPL} \textbf{109} 40009
\bibitem{shengsr} Sheng Y B and  Zhou L 2015 \emph{Sci. Rep.} \textbf{5}  7815

\bibitem{BSA} Lee S W, Park K, Ralph T C and Jeong H 2015 \emph{Phys. Rev. Lett.} \textbf{114} 113603

\bibitem{cavity}  An J H, Feng M and Oh C H 2009 \emph{Phys. Rev. A} \textbf{79} 032303

\bibitem{gate} Chen Q and Feng M 2009 \emph{Phys. Rev. A} \textbf{79} 064304
\bibitem{gate1}Chen Q and Feng M 2010 \emph{Phys. Rev. A} \textbf{82} 052329

\bibitem{cavity1}  Chen J J,  An J H, Feng M and Liu G  2010 \emph{J. Phys. B} \textbf{43} 095505

\bibitem{zhouatom} Zhou L and  Sheng Y B 2014 \emph{Phys. Rev. A} \textbf{90} 024301

\bibitem{zhouentropy} Zhou L and  Sheng Y B 2015 \emph{Entropy} \textbf{17} 4293

\bibitem{wei1}  Wei H R and  Deng F G 2013 \emph{Phys. Rev. A} \textbf{88} 042323
\bibitem{wei2} Wei H R and  Deng F G 2014 \emph{Opt. Exp.} \textbf{22} 593

\bibitem{zhoupra2} Zhou L and  Sheng Y B 2015 \emph{Phys. Rev. A} \textbf{92} 042314

\bibitem{cteleportationqip} Bastos W P, Cardoso W B, Avelar A T, de Almeida N G and Baseia B 2012 \emph{Quant. Inf. Process.} \textbf{11} 1867

\bibitem{Faraday} Julsgaard B, Kozhekin A and Polzik E S 2001 \emph{Nature} \textbf{413} 400
\bibitem{Pengpra} Peng Z H, Zou J, Liu X J, Xiao Y J and Kuang L M 2012 \emph{Phys. Rev. A} \textbf{86} 034305

\bibitem{Rb1} Nu$\beta$mann S, Hijlkema M, Weber B, Rohde F, Rempe G and Kuhn A 2005 \emph{Phys. Rev. Lett.} \textbf{95} 173602
\bibitem{Rb2}  Fortier K M,  Kim S Y, Gibbons M J, Ahmadi P and  Chapman M S 2007 \emph{Phys. Rev. Lett.} \textbf{98} 233601
\bibitem{Rb3} Colombe Y, Steinmetz T, Dubois G, Linke F, Hunger D and Reichel J 2007 \emph{Nature} \textbf{450} 272


\bibitem{atomefficiency}  Heine D, Rohringer W, Fischer D, Wilzbach M, Raub T, Loziczky S,  Yuan L X, Groth S, Hessmo B and Schmiedmayer J 2010 \emph{New J. Phys.} \textbf{12} 095005
\bibitem{photonefficiency}  D'Auria V,  Lee N, Amri T,  Fabre C, Laurat J 2011 \emph{Phys. Rev. Lett.} \textbf{107} 050504
\bibitem{photonefficiency1}  Lita A E,  Miller A J and  Nam S W 2008 \emph{Opt. Exp.} \textbf{16} 3032

\end{thebibliography}
\end{document}